\documentstyle[aps,prl,twocolumn,floats]{revtex}     
\begin{document}
\draft
\wideabs{
\title{Building a Nest at Tree Level:  \\
    Classical Metastability and Non-Trivial Vacuum Structure\\ 
    in Supersymmetric Field Theories}
\author{Keith R. Dienes \,and\, Brooks Thomas}
\address{Department of Physics, University of Arizona, Tucson, AZ  85721 USA}
\address{E-mail addresses:  ~{\tt dienes,brooks@physics.arizona.edu}}
\maketitle
\begin{abstract}
   It is becoming increasingly clear that metastable vacua may play a prominent role 
in supersymmetry-breaking.  To date, however, this idea has been realized only
in models where non-perturbative dynamics complicates the analysis of metastability. 
In this paper, we present a simple construction in which metastable vacua occur 
classically, i.e., at tree-level, and in which supersymmetry-breaking 
is sourced by both $D$-terms and $F$-terms.  
All relevant dynamics is perturbative, and hence calculations of vacuum energies and 
lifetimes can be performed explicitly.  Moreover, we find that our construction can even give rise to 
multiple non-supersymmetric vacua which are degenerate.
The non-trivial vacuum structure of such models therefore suggests that they 
can provide a rich arena for future studies of vacuum metastability in
supersymmetric field theories.
Our results may also have  
important consequences for $Z'$ phenomenology and the string landscape.   
\end{abstract}
%  \pacs{12.60.Jv,11.27.+d,14.70.Pw,11.25.Mj}
\bigskip
\bigskip
          }

%========================================================================
%          KEYSROKE-SAVING MACROS, nothing complicated 
%========================================================================
\newcommand{\newc}{\newcommand}
\newc{\gsim}{\lower.7ex\hbox{$\;\stackrel{\textstyle>}{\sim}\;$}}
\newc{\lsim}{\lower.7ex\hbox{$\;\stackrel{\textstyle<}{\sim}\;$}}

\def\beq{\begin{equation}}
\def\eeq{\end{equation}}
\def\beqn{\begin{eqnarray}}
\def\eeqn{\end{eqnarray}}
\def\calM{{\cal M}}
\def\calV{{\cal V}}
\def\calF{{\cal F}}
\def\half{{\textstyle{1\over 2}}}
\def\quarter{{\textstyle{1\over 4}}}
\def\ie{{\it i.e.}\/}
\def\eg{{\it e.g.}\/}
\def\etc{{\it etc}.\/}

%     The following macros are to create the "blackboard bold"
%     characters for "R" (set of real numbers),
%     "C" (set of complex numbers), and "Q" (set of rational numbers).

\def\inbar{\,\vrule height1.5ex width.4pt depth0pt}
\def\IR{\relax{\rm I\kern-.18em R}}
 \font\cmss=cmss10 \font\cmsss=cmss10 at 7pt
\def\IQ{\relax{\rm I\kern-.18em Q}}
\def\IZ{\relax\ifmmode\mathchoice
 {\hbox{\cmss Z\kern-.4em Z}}{\hbox{\cmss Z\kern-.4em Z}}
 {\lower.9pt\hbox{\cmsss Z\kern-.4em Z}}
 {\lower1.2pt\hbox{\cmsss Z\kern-.4em Z}}\else{\cmss Z\kern-.4em Z}\fi}
%========================================================================

\input epsf

%========================================================================
%========================================================================
%               MAIN TEXT BEGINS HERE
%========================================================================

%========================================================================
\section{Introduction}

It is no exaggeration to say that 
the physical properties of most theoretical models of particle physics
are in large part determined by the structure of their vacua.
This vacuum structure 
often determines, for example, whether the apparent symmetries of a 
model remain manifest or are spontaneously broken, 
and this in turn governs much of the resulting phenomenology.
However, it is also of considerable importance
to understand whether the vacuum structure
of a model consists of a single, unique ground state, or whether there might also exist
one or more metastable states.
Even when the true ground state 
preserves the apparent symmetries of a model,
the physical properties associated with the metastable vacua
can often differ markedly from those of the ground state. 
In such situations, the resulting phenomenology of the model
might be determined by the 
properties of a metastable vacuum rather than by those
of the true ground state.

This last question takes on a particular urgency
within the context of models exhibiting spontaneous (and indeed dynamical~\cite{WittenDSB}) 
supersymmetry-breaking, for it is well-known (see, e.g., 
Refs.~\cite{EllisEtAl,DineNelsonMetastable,PreISSDine,Dimopoulos,Luty,PreISSLuty,PreISSBanks}) 
that the scalar potentials of supersymmetric field theories often 
contain not only a global minimum corresponding to the true ground state,
but also one or more additional local minima which correspond to metastable
vacua.  Indeed, a number of recent papers~\cite{ISS,ISSR} have sparked a renewed interest
in this phenomenon.  Since the decay rates of the metastable vacua are often
highly suppressed, such metastable configurations are frequently exploited to construct 
field-theoretic models of  
dynamical supersymmetry-breaking~\cite{PostISS,DineRetro,Benakli,Shih} in
situations in which the true ground state of the theory is known
to preserve supersymmetry   by virtue of Witten-index 
arguments~\cite{WittenIndex}. 
Such metastable vacuum solutions can also be realized in a string context
through explicit D-brane constructions~\cite{MetaString}.

It has even been suggested~\cite{ISSR} that
metastable supersymmetry-breaking in such models is inevitable.
Even though minor modifications (such as the
introduction of explicit $R$-symmetry-breaking operators to give mass to the otherwise massless 
$R$-axion~\cite{SeibergNelson,RAxionSol}) can modify the Witten index 
in such theories and thereby reintroduce a supersymmetric vacuum elsewhere in field space,
such modifications tend not to spoil the properties of the false vacuum.  
The phenomenon of metastable supersymmetry-breaking  
then persists intact.

Despite their ubiquity,
dealing with such metastable vacua in supersymmetry-breaking scenarios is generally not
particularly easy.  There are two primary reasons why this is the case.
First, metastable vacua are often difficult to 
locate, enumerate, or analyze in an arbitrary model --- especially one whose ultraviolet 
completion involves non-perturbative dynamics or duality arguments.  
While a wealth of tools (including the Witten index~\cite{WittenIndex}, 
the ADS criterion~\cite{ADS}, etc.)\  are 
available for probing the ground states of supersymmetric field theories, 
far fewer tools exist for analyzing metastable vacua.  
In fact, merely establishing whether such vacua even exist in a given model 
is often difficult.  

Second, however, even if one or more metastable minima are identifiable in a given model, 
it often turns out that the low-energy effective description of such a model 
will typically have the property 
that the metastable vacua and the true vacuum are separated by 
infinite distances in field space.
As a result, any low-energy effective description of the theory which is 
valid in the vicinity of a metastable vacuum generally tends to break down 
for the true ground state (and vice versa) due to the presence of strong dynamics 
or unspecified high-scale physics.  
This implies that precise calculations of quantities which depend on 
the global features of the scalar potential ---
such as the lifetime of the metastable vacuum --- are often difficult to perform. 

In this paper, we shall present a simple theoretical construction which evades these difficulties.
Specifically, we shall present a relatively straightforward supersymmetric model whose vacuum structure exhibits 
\begin{itemize}
\item a supersymmetric ground state in which $R$-symmetry is preserved;
\item a metastable state in which both supersymmetry and $R$-symmetry are broken, and whose
        gauge symmetry also differs from that of the true vacuum;  and
\item a vacuum energy barrier between the two of a sort that results in a long lifetime
      for the metastable vacuum.
\end{itemize}
Most importantly, all three of these features will be realized {\it classically}\/, i.e., at tree level, 
and no non-perturbative physics will be required in order to create either of these vacua or guarantee their stability.
Moreover, due to the perturbative nature of our theory, these features are expected to be robust against quantum corrections.    
Finally, as we shall demonstrate, the true and metastable vacua in our model are separated by only 
a finite distance in field space. 
As a result, we will be able to perform explicit calculations of physical quantities such as the 
vacuum energy, lifetime, and particle mass spectrum of the metastable vacuum.

What makes our construction unique relative to most previous discussions 
of metastable supersymmetry-breaking in the literature
is that the supersymmetry-breaking in our model is  sourced by the presence of Fayet-Iliopoulos $D$-terms~\cite{FI}.  
While $D$-term breaking normally poses phenomenological difficulties, we shall see that in our model,
this $D$-term breaking in turn sources $F$-term breaking.  This then makes possible the breaking of
a global $R$-symmetry.

We stress that it is not our aim in this paper to present a 
complete phenomenological model of metastable supersymmetry-breaking.
While this is clearly an important endeavor, our goal here is merely to develop a 
simple, field-theoretic ``kernel'' 
which might ultimately serve as the supersymmetry-breaking sector in a fully developed model.
As such, we envision that this kernel might eventually be connected to the Standard-Model
sector through a suitable messenger sector, and likewise that this kernel might have
a suitable ultraviolet completion that allows it to emerge at sufficiently low mass scales
from a fully-developed high-scale theory.

For reasons to be discussed, however,
we do believe that 
a sensible ultraviolet completion of our kernel should not be difficult to construct,
and that this kernel could therefore 
easily be used as a platform for constructing models where supersymmetry is 
broken at scales parametrically lower than the Planck or string scale.  
Moreover, as we shall discuss,
we believe that our kernel has another important property required of all 
models of metastable supersymmetry-breaking, 
namely that the properties of the metastable vacuum in our model will not be spoiled 
by the introduction of additional operators with small coefficients, even if these operators
tend to benignly produce additional supersymmetric vacua elsewhere in field space.
Consequently, we believe that a variety of phenomenological model-building requirements  ---
such as successfully coupling the model to a messenger sector, and including an explicit component to
$R$-symmetry breaking to avoid running afoul of experimental constraints on a massless $R$-axion~\cite{SeibergNelson} ---
should be easy to satisfy.

Finally, we point out that our kernel is remarkably simple, consisting of 
only two $U(1)$ gauge groups and a superpotential of the Wilson-line type.  This gives our
kernel a relevance beyond the narrow issue of supersymmetry-breaking.  For example,
structures of this sort appear naturally --- and in fact are almost unavoidable --- 
in heterotic and Type~I orientifold string models, and as such we expect our discussion of the vacuum structure
of our models to have a relevance in string theory that transcends their potential use as kernels 
in supersymmetry-breaking model-building.  One possible implication is that the presence of 
long-lived metastable vacua in such models should have a significant effect 
on overall landscape statistics~\cite{douglas}, which could in turn affect the methodologies
that have been employed for statistical samplings of explicit heterotic~\cite{hetlandscape} 
and Type~I~\cite{TypeIlandscape} string vacua.  
More phenomenologically, the non-trivial vacuum structure of 
the multiple-$U(1)$ theories we discuss has the possibility to lead to new observable signatures for $Z'$ physics.

%========================================================================
\section{The Model\label{sec:Model}}

Our model is surprisingly simple.  Working in terms of ${\cal N}{=}1$ supersymmetry, our
model consists of two $U(1)$ gauge groups denoted $U(1)_a$ and $U(1)_b$, with respective gauge
couplings $g_{a,b}$ and Fayet-Iliopoulos terms
$\xi_{a,b}$, and five chiral superfields with charges as shown
in Table~\ref{tab:chgs}.  Given this configuration, the most 
general renormalizable superpotential is   
given by
\begin{equation}
             W~=~\lambda\,\Phi_1\Phi_2\Phi_3 \, +\, m\, \Phi_4\Phi_5~.
\label{eq:W}
\end{equation}
Indeed, the $(\Phi_1,\Phi_2,\Phi_3)$  ``core'' of this model is nothing but a two-site linear moose of the 
sort that has been discussed intensely in the literature 
in connection with deconstruction~\cite{decon}, flux compactifications
of string theory~\cite{flux}, and even toy field-theoretic models of the string landscape~\cite{DDG}. 
To this core, we have then simply added a vector-like
pair of fields $(\Phi_4,\Phi_5)$ charged under both $U(1)_a$ and $U(1)_b$ with a supersymmetric 
mass $m$.  
As discussed in Ref.~\cite{DudasDecon},
the mixed anomalies implicit in this configuration
may be cancelled by adding moduli fields $S_n$ that couple to the gauge fields
through a term of the form $\sum_nS_nW_n^{\alpha}W_{n\alpha}$, where $n=a,b$.

%==================== TABLE BEGINS HERE ====================================
\begin{table}[t!]
\begin{center}
\begin{tabular}{c|ccc}
      Field & $U(1)_a$ & $U(1)_b$ & $U(1)_R$ \\ 
       \hline
     $\Phi_{1}$ & $-1$ & $0$ & $2/3$ \\
     $\Phi_{2}$ & $+1$ & $-1$  & $2/3$ \\
     $\Phi_{3}$ & $0$ & $+1$ & $2/3$ \\
     $\Phi_{4}$ & $+1$ & $+1$ & $1$  \\
     $\Phi_{5}$ & $-1$ & $-1$ & $1$ \\ 
\end{tabular}
\end{center}
\caption{The field content and charge assignments for the model under consideration.}
\label{tab:chgs}
\end{table}
%==================== TABLE BEGINS HERE ====================================

Such a model has a rich phenomenology.  For simplicity, we shall consider the gauge
couplings to 
be equal, $g_a=g_b\equiv g$, and we shall set $g=1$ for convenience.
Our model is then defined in terms of the four
remaining parameters $(\lambda, m, \xi_a,\xi_b)$. 
However, for the purpose of analyzing the
metastability inherent in this model, 
we will begin by considering by considering
the simple parameter choice $(\lambda,m,\xi_a,\xi_b)=(1,1,5,0)$,
where we have rescaled all dimensionful parameters with appropriate
powers of an overall, unspecified mass scale.
Other choices for the parameters shall be discussed below.

In general, the scalar potential for a supersymmetric gauge theory coupled to matter includes
both $D$-term and $F$-term contributions and can be written in the form
\begin{equation}
      V~=~ \half \sum_a g_a^2 D_a^2 + \sum_i  |F_i|^2~,
\end{equation}
where
\begin{equation}
  D_a = \xi_a+\sum_i q_i^{(a)}|\phi_i|^2~,~~~~~~
  F_i =-\frac{\partial W^{\ast}}{\partial\phi^{\ast}_i}~.
\end{equation}
In the model under consideration here, the scalar potential, including both $D$-term and $F$-term 
contributions, is given by
\begin{eqnarray}
  V ~&=&~\lambda^2(|\phi_1|^2|\phi_2|^2+|\phi_1|^2|\phi_3|^2+|\phi_2|^2|\phi_3|^2)\nonumber\\
  &&~~~+m^2(|\phi_4|^2+|\phi_5|^2)\nonumber\\
  &&~~~+\half g_a^2(\xi_a-|\phi_1|^2+|\phi_2|^2+|\phi_4|^2-|\phi_5|^2)^2\nonumber\\
  &&~~~+\half g_b^2(\xi_b-|\phi_2|^2+|\phi_3|^2+|\phi_4|^2-|\phi_5|^2)^2~.
\label{eq:V}
\end{eqnarray}
Note that $V$ is a function only of absolute squares of fields; 
we can therefore take the vacuum expectation values of all the $\phi_i$
to be real and positive-semidefinite without loss of generality.  
For a given choice of the model parameters $(\lambda,m,\xi_a,\xi_b)$,
the extrema of the scalar potential
can then be obtained by solving the coupled simultaneous equations
\begin{equation}
          \frac{\partial V}{\partial\phi_i}=0~~~~~(i=1,\ldots,5)~.
\label{eq:ExistEqs}
\end{equation}
However, a solution is a local minimum only if the eigenvalues of the $10\times 10$ mass matrix
\begin{equation}
  {\mathcal M}^2~\equiv~ \pmatrix{
  \frac{\partial^2 V}{\partial\phi_i^\ast \partial\phi_j} &
  \frac{\partial^2 V}{\partial\phi_i^{\ast}\partial\phi_j^\ast} \cr
  \frac{\partial^2 V}{\partial\phi_i\partial\phi_j} &
  \frac{\partial^2 V}{\partial\phi_i\partial\phi_j^\ast} \cr}
\label{matrix}
\end{equation}
are all non-negative and the number of zero eigenvalues is precisely equal to the
number of Goldstone bosons eaten by the massive gauge fields.
(Indeed, additional
zeroes would indicate the presence of classical flat directions.)
In what follows, however, we will use the term ``vacuum'' loosely to refer to 
any extremum of the potential and employ adjectives such as ``stable'' and ``unstable''
to distinguish the eigenvalues of the mass matrix.        
Of course, a ``metastable'' vacuum exists only when two or more vacua exist and
are stable according to the above definitions;  all but the vacuum with lowest energy
are considered metastable.
Finally, we shall refer to a given vacuum as potentially breaking $R$-symmetry
if its associated field configuration implies that any of the $F$-terms receive 
a non-zero VEV.

%=========================== TABLE BEGINS HERE ================================================
 \begin{table*}[thb!]
\begin{center}
  \begin{tabular}{ccccccc}
     Label & $(v_1,v_3,v_5)$            &  $V$      & Stability  & SUSY & $R$-symmetry & Gauge Group  \\ 
           \hline
     A & $(\sqrt{5},0,0)$               & $0$       & Stable     & Yes & Yes & $U(1)_b$  \\
     B & $(0,2,2)$                      & $9/2$     & Metastable & No  & No  & None  \\
     C & $(\sqrt{3}/2, \sqrt{7}/2,\sqrt{5/2})$ & $45/8$    & Unstable   & No  & No  & None  \\ 
     D & $(0,0,\sqrt{2})$                      & $17/2$     & Unstable   & No  & No  & $U(1)_{(a-b)}$  \\
     E & $(0,0,0)$                      & $25/2$     & Unstable   & No  & Yes & $U(1)_a\times U(1)_b$ \\
   \end{tabular}
\end{center}
 \caption{The complete classical vacuum structure of the model in Eq.~(\ref{eq:W})     
       with $(\lambda,m,\xi_a,\xi_b)=(1,1,5,0)$ and $g=1$.
       For each of the five extrema (labelled A through E) of the scalar potential 
       in Eq.~(\ref{eq:V}), we have listed
       the corresponding VEVs $(v_1,v_3,v_5)$, the value of the scalar potential,
       and the stability properties of that extremum.  We have also indicated whether supersymmetry
       and $R$-symmetry are broken or unbroken at that extremum,
       along with the surviving (unbroken) gauge group.
       Each of these solutions has $v_2^2=v_4^2=0$;  likewise,
        all dimensionful quantities are
       quoted in terms of the overall arbitrary mass scale associated with our model.
       Note that the unstable extrema $D$ and $E$ are local maxima in the 
       restricted three-dimensional VEV subspace $(v_1,v_3,v_5)$,
       but are actually saddle points in the full five-dimensional
       VEV space $(v_1,v_2,v_3,v_4,v_5)$.}
\label{tab:solsprops}
 \end{table*}
%=========================== TABLE BEGINS HERE ================================================

%================== FIGURE ============================================
\begin{figure*}[thb!]
\centerline{
   \epsfxsize 3.9 truein \epsfbox {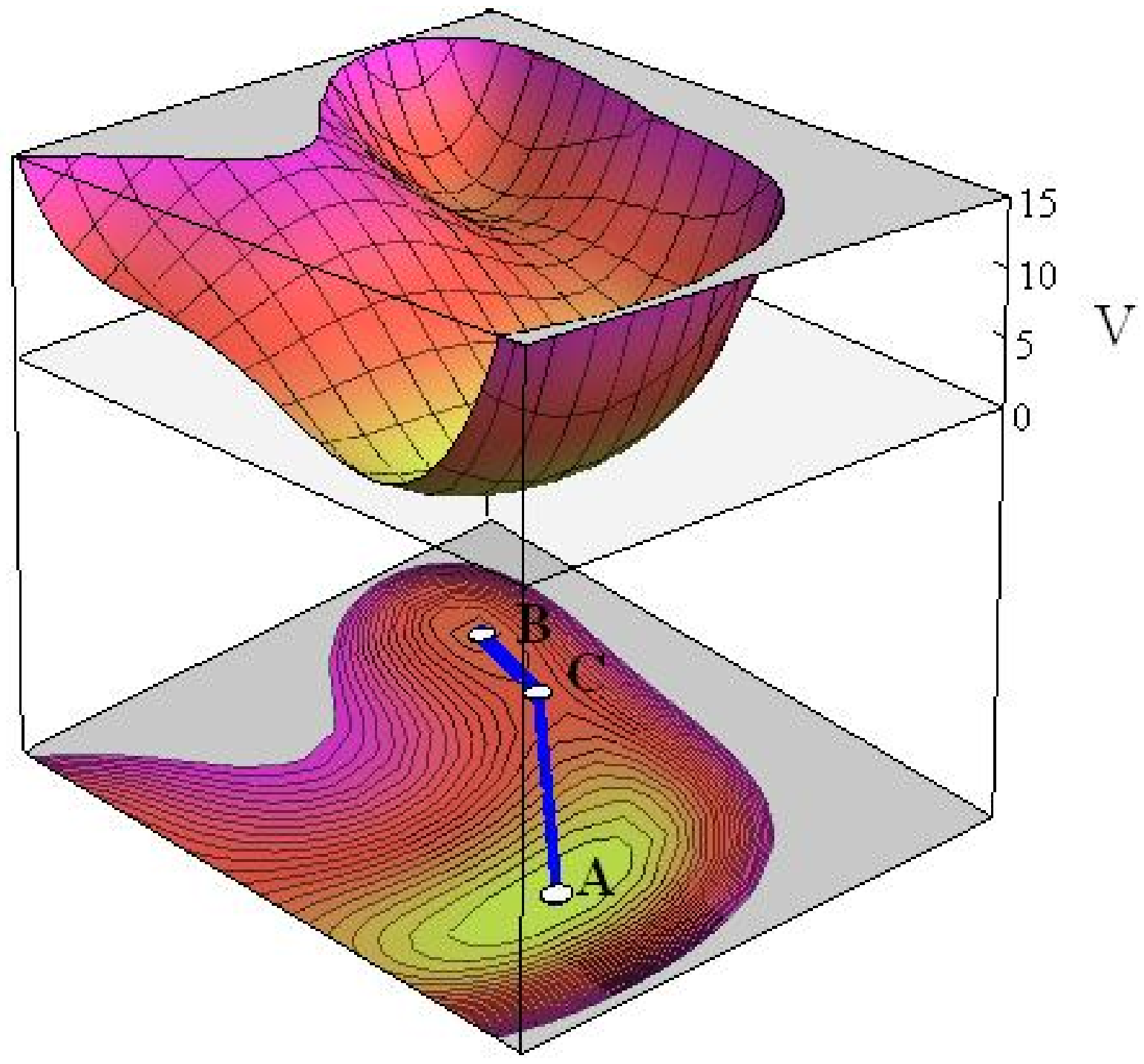} 
   \epsfxsize 3.5 truein \epsfbox {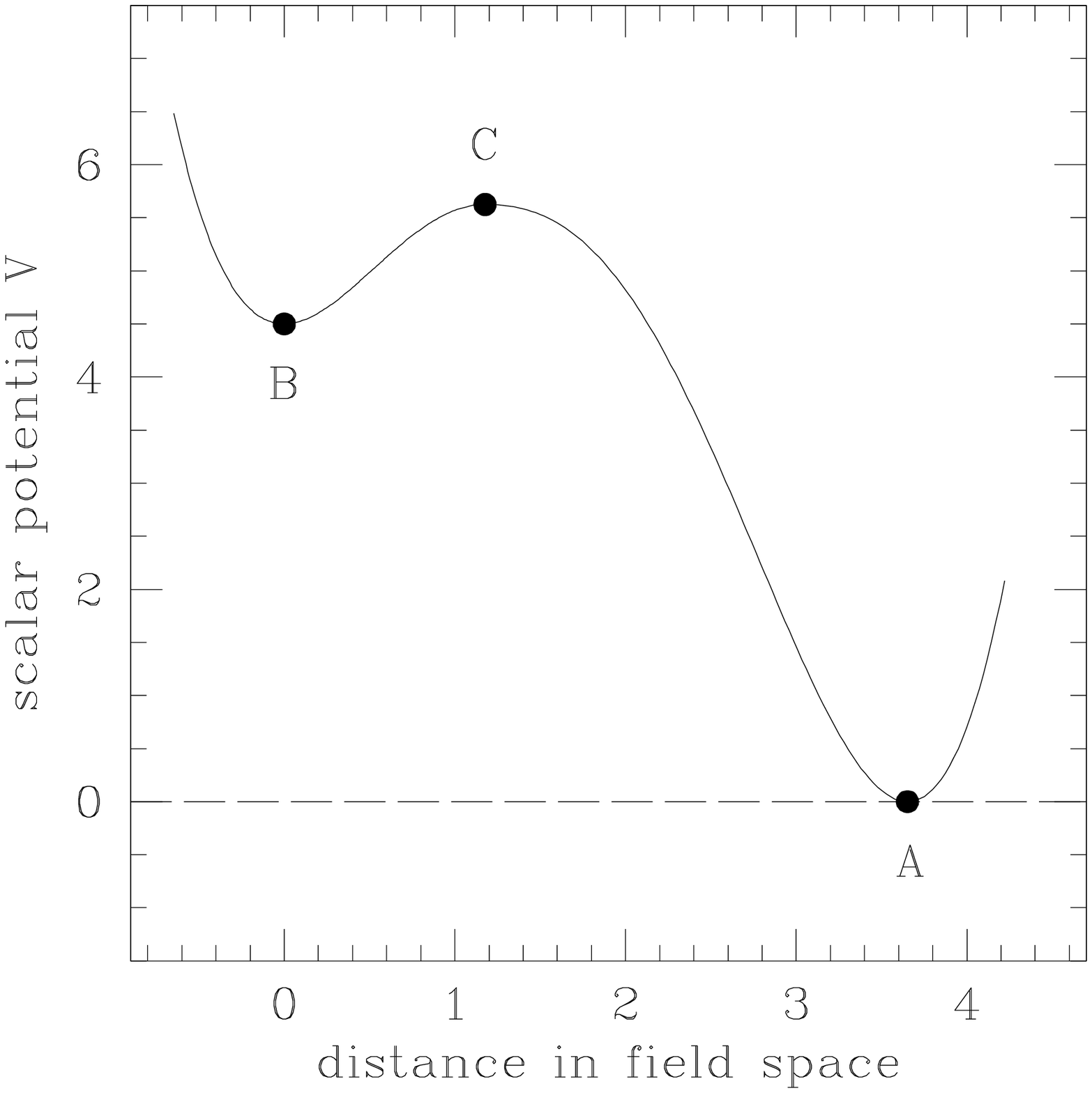} 
 }
\smallskip
\caption{The supersymmetric ground state of our model and the metastable ``nest'' above it.  
           {\it Left figure}\/:  
      A surface plot of the scalar potential $V$ evaluated on
       the unique two-dimensional plane within the three-dimensional
      $(v_1,v_3,v_5)$ field space which simultaneously contains 
      the true vacuum solution~A, 
     the metastable vacuum solution~B, 
          and the saddle-point solution~C between them.
      Projected below the surface plot is a contour plot for $V$,
      showing the shortest path (blue) in field space connecting these three
      solutions.
      {\it Right figure}\/:
       The scalar potential $V$ evaluated along this shortest path.
     Field-space distances are quoted relative to the 
        metastable vacuum~B along this path, and all units are in terms
        of the overall unspecified mass scale associated with our model.
      Note that the field-space distance between the metastable
        and true ground states is finite and $\sim{\cal O}(1)$;
      nevertheless, we shall see in Sect.~IV that the lifetime
      of this metastable vacuum exceeds the age of the universe
      even when this unspecified mass scale is taken to be as high as the Planck scale.} 
\label{Vplotfig}
\end{figure*}  
%========================================================================

We claim that the model specified in Eqs.~(\ref{eq:W}) and (\ref{eq:V}) has all the 
desired features outlined in Sect.~I.~  In particular, the
vacuum structure of the model can be explicitly described as follows:
\begin{itemize}
\item First, there exists a stable, supersymmetric ground state, henceforth denoted `A', with 
          $V=0$ and unbroken $R$-symmetry in which only one field ($\phi_1$) receives a non-zero 
          VEV.  As a result, the gauge group $U(1)_b$ is preserved in this vacuum state.
\item Second, there exists a metastable vacuum with $V=9/2$, henceforth denoted `B', in 
      which both supersymmetry and $R$-symmetry are 
          broken and in which two fields ($\phi_3$ and $\phi_5$) receive non-zero VEVs.  
          The gauge group is entirely broken in this metastable vacuum.
\item Finally, there exist {\it three}\/ additional unstable solutions 
           to Eq.~(\ref{eq:ExistEqs}), henceforth denoted `C', `D', and `E', 
          with even higher vacuum energies.  
         One preserves both $U(1)$'s, one preserves a linear combination of 
           $U(1)_a$ and $U(1)_b$, and one breaks the gauge symmetry altogether,
          leaving behind a heavy pseudoscalar in the process.  
           All three break supersymmetry, but only two break $R$-symmetry.
\end{itemize}
A complete listing of the vacuum structure of this model is given in Table~\ref{tab:solsprops}.  
Fig.~\ref{Vplotfig} shows the behavior of the scalar potential $V$ evaluated between
the metastable vacuum solution~B (our tree-level ``nest''), 
the true vacuum solution~A (the ``ground'' state),
and the saddle-point solution~C which connects them.

%========================================================================
\section{Exploring the Parameter Space\label{sec:ParamSpace}}

Thus far, we have presented a model in which there exists both a supersymmetric
ground state and a non-supersymmetric metastable state at tree level.
This is the model with the particular choices $(\lambda,m,\xi_a,\xi_b)=(1,1,5,0)$
and $g=1$.
In this section, we seek to understand the behavior of this model as all of 
these parameters are varied, with an eye towards determining the conditions
under which a non-supersymmetric metastable vacuum emerges relative to a
supersymmetric ground state.

%================== FIGURE ============================================
\begin{figure*}[ht]
\centerline{
   \epsfxsize 2.3 truein \epsfbox {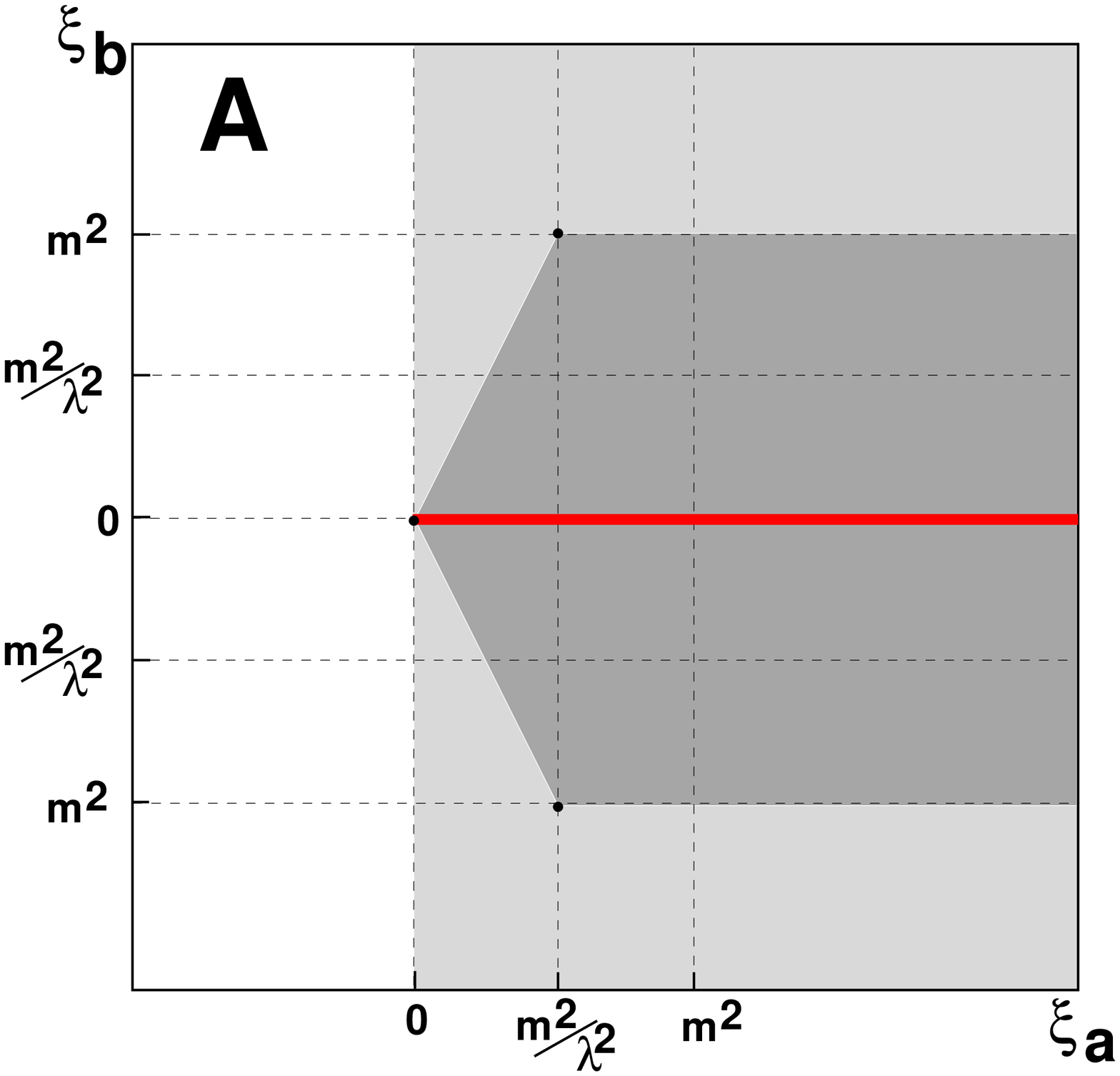} 
   \epsfxsize 2.3 truein \epsfbox {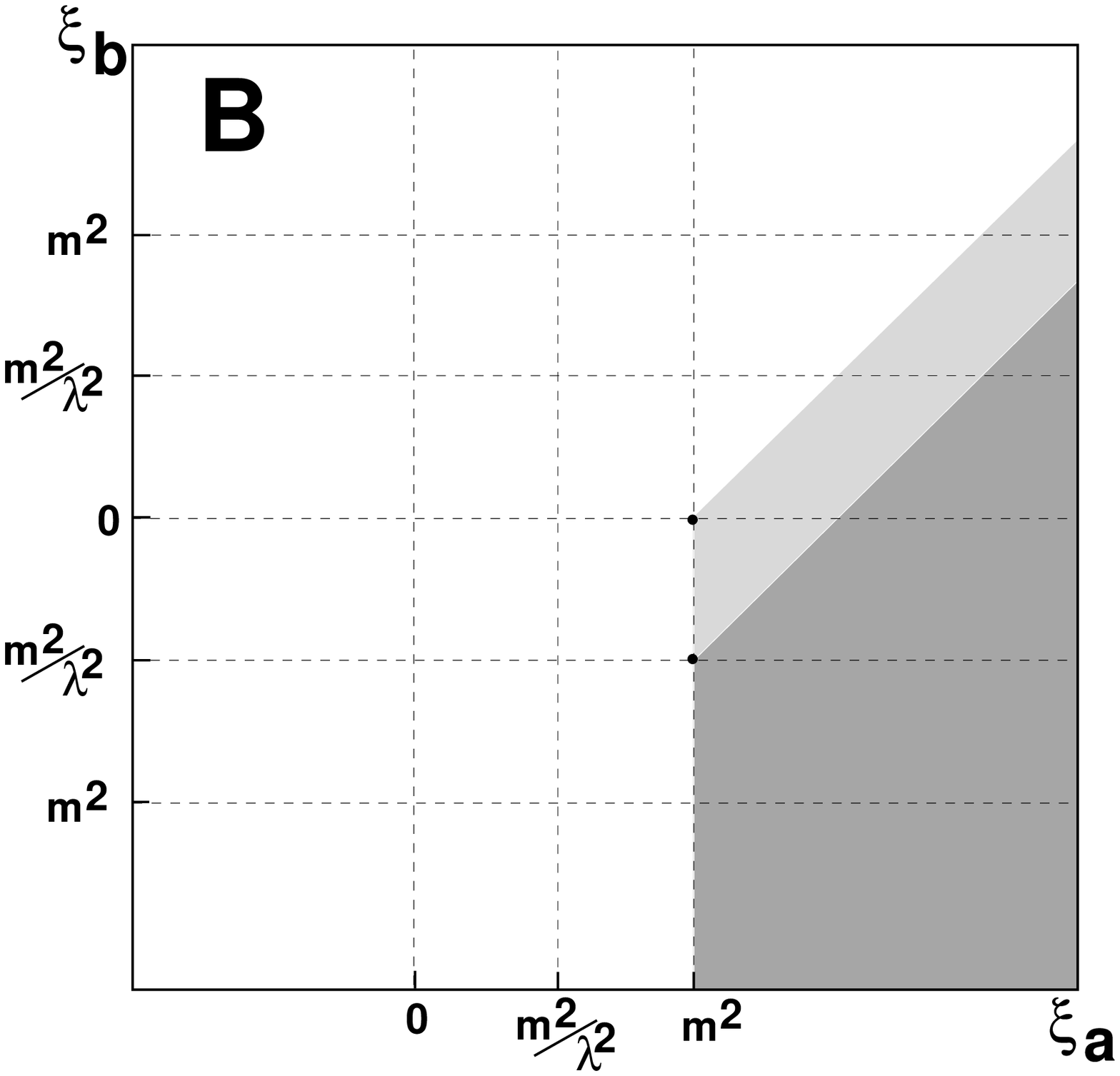}} 
\centerline{
   \epsfxsize 2.3 truein \epsfbox {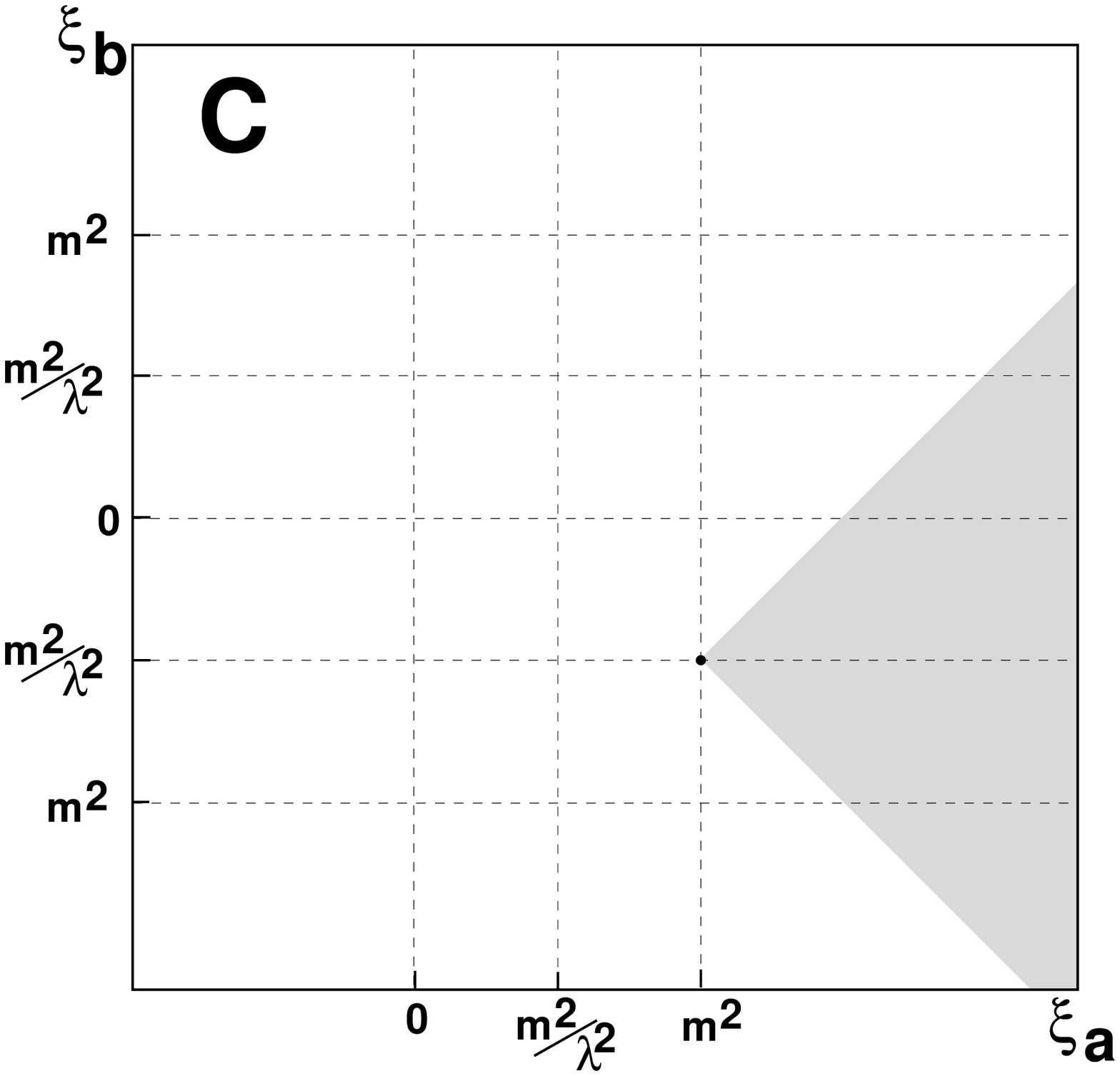} 
   \epsfxsize 2.3 truein \epsfbox {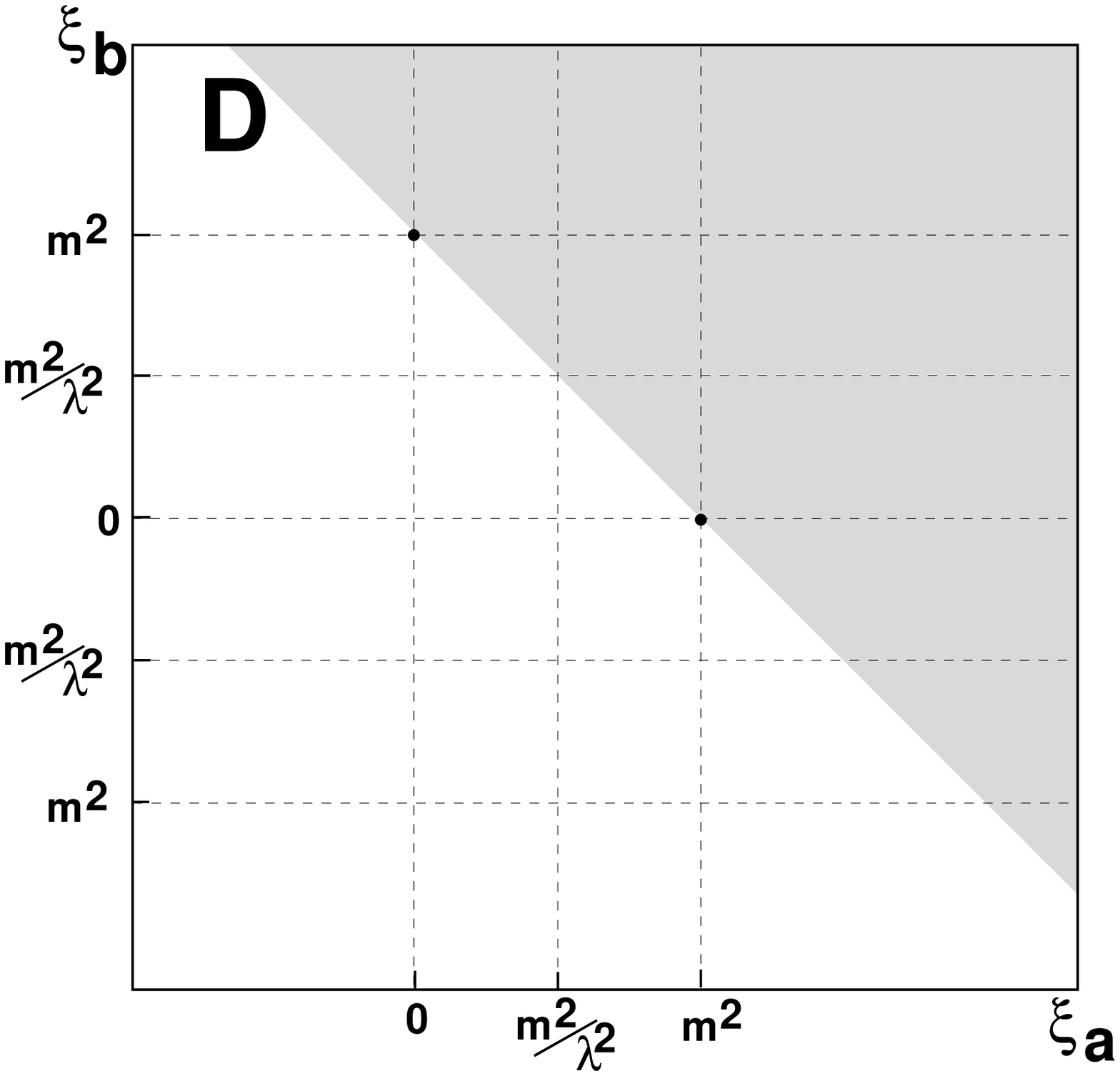} 
   \epsfxsize 2.3 truein \epsfbox {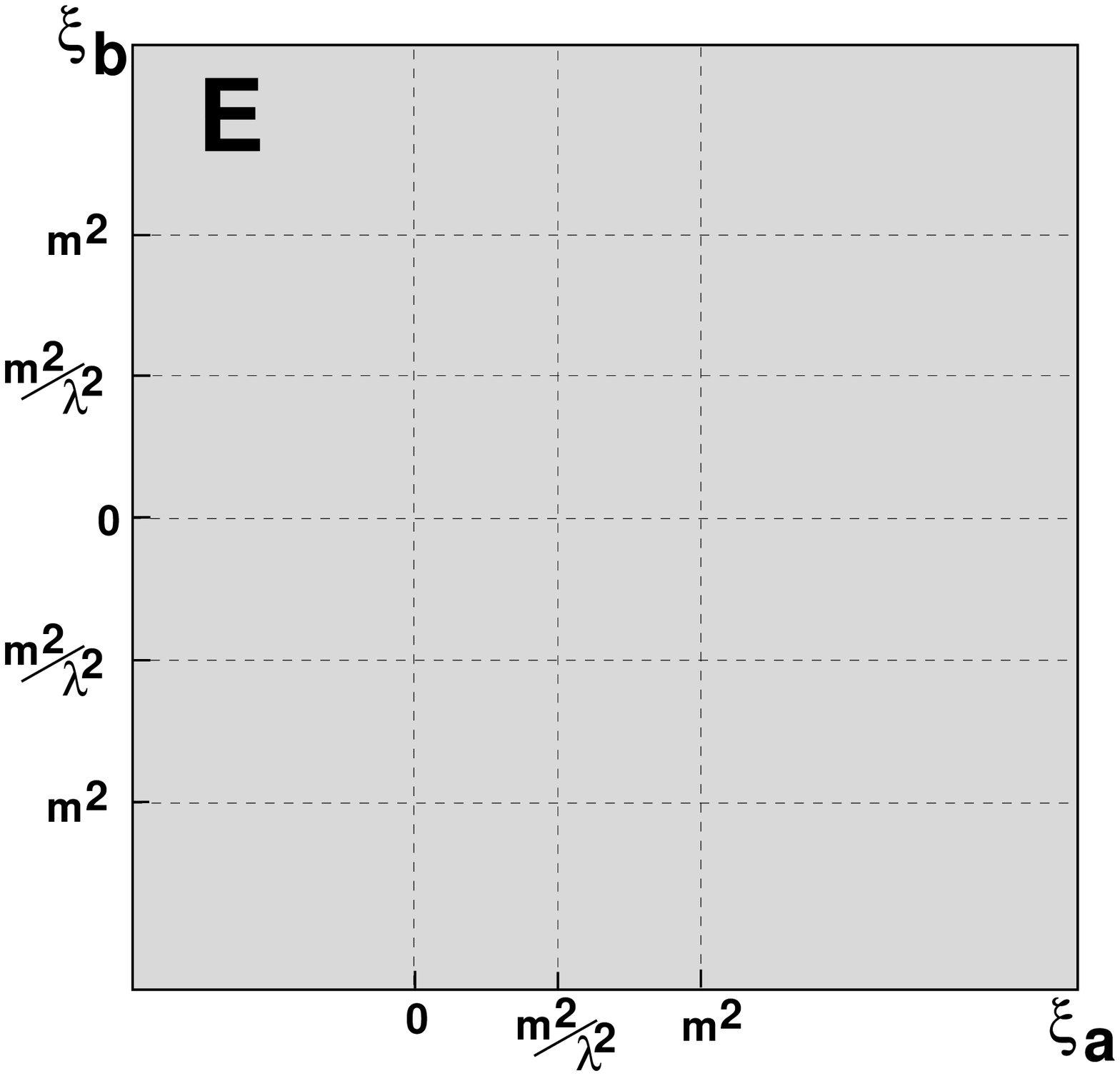}} 
\smallskip
\caption{Within the two-dimensional $(\xi_a,\xi_b)$ plane, we 
         have indicated the regions in which Solutions~A through E 
         exist (light shading) or both exist and are stable (darker shading).  
         In drawing these regions, we have implicitly assumed $\lambda^2>1$ 
         when laying out the coordinates on each axis, but the results 
         shown are general and apply even when $\lambda^2\leq 1$, with
         the shaded regions deforming accordingly. 
         Note that the solid horizontal line (red) within Solution~A indicates
         a line of vacua with unbroken supersymmetry.
      }
\label{fig:Regions}
\end{figure*}
%================== END OF INSERTED FIGURE ============================

Even though our model contains five parameters $(g,\lambda,m,\xi_a,\xi_b)$,
we can always rescale our $(\lambda,m,\xi_a,\xi_b)$ parameters so as to absorb
the gauge coupling $g$ completely and without loss of generality.  To do this, we 
can  simply rewrite our expressions in terms of the new
variables 
$(\tilde \lambda,\tilde m,\tilde \xi_{a,b},\tilde v_i)\equiv
(\lambda/g,m/\sqrt{g},g\xi_{a,b},\sqrt{g}v_i)$,
and then drop the tildes.
We shall therefore use these rescaled variables in what follows. 

It is straightforward to generalize our Solutions~A through E for arbitrary
$(\lambda,m,\xi_a,\xi_b)$, obtaining
\begin{eqnarray}
  A:~~~   && v_1^2=\xi_a,~~~v_3^2=0,~~~v_5^2=0 \nonumber\\
  B:~~~   && v_1^2=0,~~~v_3^2=\xi_a-\xi_b-m^2,~~~v_5^2=\xi_a-m^2 \nonumber\\
  C:~~~   && \cases{
            \displaystyle v_{1,3}^2=\frac{\xi_a-\xi_b}{2(\lambda^2+1)}\mp\frac{m^2}{2\lambda^2}~, & ~\cr 
            \displaystyle v_5^2=\frac{1}{2}\left[\xi_a+\xi_b+\frac{m^2}{\lambda^2}(1-\lambda^2)\right]~.& ~\cr}\nonumber\\
  D:~~~   && v_1^2=v_3^2=0~,~~~ v_5^2 = \half(\xi_a+\xi_b-m^2)\nonumber\\
  E:~~~   && v_1^2=v_3^2=v_5^2=0~ 
\label{solns}
\end{eqnarray}
where we continue to have $v_2^2=v_4^2=0$ for each solution.
Substituting these results into Eq.~(\ref{eq:V}),
we then obtain the vacuum energies for each of these field configurations:
\begin{eqnarray}
  V_A ~&=&~ \half \xi_b^2 ~,~~~~
  V_B ~=~  m^2\xi_a  - \half {m^4} \nonumber\\
  V_C ~&=&~ \frac{1}{4 \lambda^2 (1 + \lambda^2)}\,\bigl\lbrack
        \lambda^4 (\xi_a - \xi_b)^2\nonumber\\
        &&~~~ + 2 m^2 \lambda^2 (1 + \lambda^2) (\xi_a + \xi_b)
            +m^4 (1-\lambda^4)\bigr\rbrack~\nonumber\\
  V_D ~&=&~ \quarter(\xi_a-\xi_b)^2 + \half m^2 (\xi_a+\xi_b) - \quarter m^4 \nonumber\\
  V_E ~&=&~ \half(\xi_a^2+\xi_b^2)~. 
\label{Vs}
\end{eqnarray}   
Clearly, these solutions exist only for those combinations 
of parameters $(\lambda,m,\xi_a,\xi_b)$
for which these expressions for the $v_i^2$ are non-negative.
For example, we see from Eq.~(\ref{solns}) that Solution~A does not exist as 
a solution to the equations in Eq.~(\ref{eq:ExistEqs})
unless $\xi_a\geq 0$.

However, even when these solutions exist, 
we cannot easily determine from Eqs.~(\ref{solns}) and (\ref{Vs})
whether they correspond to stable or unstable extrema
of the potential.  For this, we must examine the eigenvalues of the 
mass matrix in Eq.~(\ref{matrix}) and demand that they satisfy the conditions
discussed below Eq.~(\ref{matrix}).
It is this which is the subtle part of the analysis,
and the results are shown in Fig.~\ref{fig:Regions}.
For any arbitrary values of $\lambda$ and $m$, we have
illustrated in Fig.~\ref{fig:Regions}
those regions in the two-dimensional $(\xi_a,\xi_b)$ plane 
for which each of our solutions not only exists but is also stable.
Such $(\xi_a,\xi_b)$ plots are similar to those originally drawn in Ref.~\cite{DDG},
and from these plots it is relatively straightforward 
to see how these regions deform as $\lambda$ and $m$ are varied.

Given these results, we can proceed to determine the general conditions
on our model parameters $(\lambda,m,\xi_a,\xi_b)$ for which 
various types of metastability emerge.
First, we observe from Fig.~\ref{fig:Regions}
that only for $\xi_b=0$ does our Solution~A
correspond to a supersymmetric vacuum;  for all other
$|\xi_b|< m^2$ this vacuum state is non-supersymmetric.
Thus $\xi_b=0$ is a necessary condition for having a supersymmetric
ground state.  We also observe from Fig.~\ref{fig:Regions} that 
our theory will also have a non-supersymmetric 
metastable vacuum as long as $\xi_a > (1+1/\lambda^2)m^2$.
We thus conclude that both
a supersymmetric ground state and a non-supersymmetric metastable
state will simultaneously be achieved in our model
for all $(\lambda,m,\xi_a,\xi_b)$ satisfying
\beq
    \cases{  \xi_a > (1+1/\lambda^2)m^2~&~\cr
            \xi_b=0~. & ~\cr }
\label{susycase}
\eeq

Of course, it is not absolutely necessary that our true ground state
have unbroken supersymmetry, especially if our main interest is
in the existence of the metastable state.  If we are willing 
to relax this requirement,
then we can easily realize the more general situation in which
we have two ground states, one true and one metastable,
both of which are non-supersymmetric.
Given the results in Fig.~\ref{fig:Regions}, we see that this
occurs for all $(\lambda,m,\xi_a,\xi_b)$ satisfying
the simultaneous constraints
\beq
    \cases{  
        \xi_a-\xi_b > (1+1/\lambda^2)m^2 &~\cr
        |\xi_b| < m^2  &~\cr
        \xi_a > {\rm max} (m^2, m^2/\lambda^2)~.&~\cr}
\label{nonsusycase}
\eeq
Note that the third of these constraints becomes redundant
with the first two if $\lambda^2\leq 1$.  
Also note that we can make our 
true ground state as ``approximately'' supersymmetric as we wish
by taking $|\xi_b|$ sufficiently small;  in the limit
$|\xi_b|\to  0$, the constraints in Eq.~(\ref{nonsusycase}) then
reduce to those in Eq.~(\ref{susycase}).
Finally, we observe that the conditions in Eq.~(\ref{nonsusycase})
always guarantee that $V_A < V_B$ despite the fact that both
vacua are non-supersymmetric.  Thus Solution~B continues
to correspond to the metastable vacuum.

Given the general results sketched in Fig.~\ref{fig:Regions},
it is also possible to understand how our model achieves its
metastability.  Let us begin
by considering the limit in which $m^2/\xi_a, m^2/\xi_b\to\infty$.
At low energies, this limiting case reduces to the model considered in Ref.~\cite{DDG}: 
$\Phi_4$ and $\Phi_5$ become infinitely massive and decouple, 
leaving only the two-site orbifold moose diagram.
Indeed, as $m^2/\xi_a,m^2/\xi_b \to \infty$, 
the regions corresponding to Solutions~B, C, and D recede off
the $(\xi_a,\xi_b)$ plane entirely, and the region of stability for Solution~A 
expands to fill a triangular ``pie-slice''~\cite{DDG} of angle $2\theta_\lambda$, where 
$\tan\theta_\lambda\equiv \lambda^2$.
Of course, our $\xi_b=0$ supersymmetric ground states within Solution~A persist without
alteration.  However, no metastable vacua are present above them.

Therefore, in order to obtain additional metastable vacua,
it is necessary to go beyond the model in Ref.~\cite{DDG} by introducing the massive, 
vector-like pair of fields
that appear in Eq.~(\ref{eq:W}).  Indeed, as $m$ decreases from infinity,
we see that our metastable vacua~B begin to re-enter the picture and eventually
overlap the supersymmetric ground states within Solution~A.  
Thus, while it is the Wilson-line term $\lambda \Phi_1\Phi_2\Phi_3$ within the 
superpotential in Eq.~(\ref{eq:W}) that produces our supersymmetric ground state,
it is the $m\Phi_4\Phi_5$ mass term in Eq.~(\ref{eq:W}) that introduces the metastable vacuum.
It is truly remarkable
that the introduction of the mass term does not disturb the original 
supersymmetric solution, and that these two solutions can peaceably co-exist
within the same regions of parameter space.  Our experience with other models has shown
that this is not often the case.

Given the critical role played by the term $m\Phi_4\Phi_5$ in producing the metastable vacuum,
one might wonder what happens when further vector-like pairs of chiral superfields 
with supersymmetric masses are introduced into our model.  
For example, we might consider an extended 
superpotential of the form
\beq
         W ~=~ \lambda \Phi_1\Phi_2\Phi_3 + m_1\Phi_4\Phi_5  + m_2\Phi_6\Phi_7 + m_3\Phi_8\Phi_9 + ...
\label{extravecs}
\eeq
where $(\Phi_6,\Phi_7)$, $(\Phi_8,\Phi_9)$, etc.,  have
the same $U(1)$ and $R$-charges as $(\Phi_4,\Phi_5)$. 
(Note that $W$ continues to be the most general renormalizable superpotential one can write
in such situations, since potential cross-couplings such as $\Phi_4\Phi_7$ can always
be removed without
loss of generality by diagonalizing the corresponding effective mass matrix.)
At first glance, it might seem that the introduction of such extra vector-like fields
would introduce a whole tower of metastable vacua, with
$m_i$ governing their spectrum of vacuum energies.
However, it turns out that {\it only the  lightest}\/ pair of fields introduces
a metastable vacuum, 
while the remaining fields produce extraneous extrema which are all unstable.    
%  Yo!  Wake up.
Moreover,
this metastable vacuum is identical to Solution~B above, with the corresponding
mass $m_i$ playing the role formerly played by $m$. 

This is a very important observation, indicating that it is possible to couple arbitrary numbers
of additional vector-like pairs of chiral fields into our model,
all while preserving the metastability we have already observed.
This also demonstrates that the mass scale $m$ associated with the metastability in our 
model
is always the {\it smallest}\/ of the supersymmetric masses in the model, 
and {\it remains so regardless
of the masses of such additional fields, no matter how large they may be}.
Thus, if such additional fields are messengers to
a separate Standard-Model sector of the theory, we see that our model can consistently
accommodate a messenger energy scale which is significantly higher than 
that associated with the supersymmetry-breaking scale of the metastable vacuum. 
This is exactly what would be required for both gravity- and gauge-mediated scenarios
of supersymmetry-breaking.

We can also calculate the particle spectra corresponding to each of the two vacuum states
in our model.
The analysis is straightforward, but may have important phenomenological implications
if our model is to be used as a potential supersymmetry-breaking sector
(or more generally as a potential hidden sector coupled to the Standard Model at an intermediate
scale).
Of course, as part of the spectrum analysis, we must take into account the effects
of both gauge-symmetry breaking and supersymmetry-breaking, when relevant. 

Restricting to the $(\lambda,m,\xi_a,\xi_b)$ region indicated in Eq.~(\ref{susycase}),
our results are as follows.
Recall that our model consists of five complex scalars $\phi_i$ ($i=1,...,5$),
five Weyl fermions $\psi_i$, two $U(1)$ gauge fields $A_{a,b}$, and
their superpartner gauginos $\lambda_{a,b}$.
For the true vacuum solution, we then find that $U(1)_b$ remains
unbroken, while $U(1)_a$ is broken spontaneously by the VEV of $\phi_1$.
The corresponding gauge boson then acquires a squared mass $M^2_{Z'}=2\xi_a$.  
As a result of the unbroken supersymmetry,
the gaugino associated with the $U(1)_b$ gauge group remains massless;  
thus Solution~A always gives rise to a massless field with odd $R$-parity
unless there are other sources of supersymmetry-breaking beyond our model.
The lightest remaining physical states in the spectrum are 
the $\phi_{4,5}$ fields, which obtain squared mass $m^2$, and their superpartners.

In the metastable vacuum, the situation is of course quite different.  
Here, we find that the non-zero expectation values for $\phi_3$ and $\phi_5$ spontaneously break  
both $U(1)_a$ and $U(1)_b$, resulting in a pair of heavy $Z'$ gauge bosons with
squared masses   
\begin{equation}
         M_{Z'}^2 ~=~ (3\pm\sqrt{5})(\xi_a-m^2)~.
\end{equation} 
A massless Goldstone fermion also appears in the spectrum (as one would expect, since supersymmetry
is spontaneously broken in this vacuum configuration), and its identity is an admixture of $\psi_4$ 
and $\lambda_a$.  All of the remaining physical particles are massive,
with the identity of the lightest scalar depending on $(\lambda,m,\xi_a)$ and consisting of either
$\phi_4$ with squared mass $2m^2$;
$\phi_1$ with squared mass $\lambda^2 (\xi_a-m^2) -m^2$;
or an admixture of $\phi_3$ and $\phi_5$ with squared mass
$(3-\sqrt{5})(\xi_a-m^2)$.
The situation for the corresponding fermions is similar, but more complicated
as the mixing between the $\psi_i$ fields and the gauginos is far less trivial.
However, we already see from these results that unlike the true vacuum, our metastable vacuum 
does not give rise to dangerous massless matter with odd $R$-parity.
This once again provides an explicit illustration of the phenomenological impact of being in
a metastable vacuum rather than the true ground state.

Up to this point, we have focused on those conditions under which our model
contains both a supersymmetric ground state and a metastable state at tree level.
We have not yet tackled the question of perturbativity, which is necessary 
not only for calculability but also for ensuring
that our tree-level solutions remain intact against radiative corrections.
However, it is easy to restore the factors of gauge coupling 
that were missing from Eq.~(\ref{susycase}), obtaining
\beq
    \cases{  \displaystyle \xi_a > \left(1+{g^2 \over \lambda^2}\right) {m^2\over g^2}~&~\cr
            \xi_b=0~. & ~\cr }
\eeq
Likewise, the condition for perturbativity is given by
\beq
              g^2 ~\ll~ 16\pi^2~.
\eeq
Thus, combining these results, we find that our model 
will not only 
function as advertised
but also be perturbative provided $\xi_b=0$ and
\beq
      {1\over 16\pi^2}~\ll~ {1\over g^2} ~<~ {\xi_a\over m^2} - {1\over \lambda^2}~.
\label{master}
\eeq
Note that our original parameter choices from Sect.~II amply satisfy these requirements.
While the general result in Eq.~(\ref{master}) technically places a lower bound on the 
gauge coupling, this lower bound can be as small as we wish provided 
$\xi_a$ is taken sufficiently high or $m$ is taken sufficiently small.
Thus, loop corrections can be arbitrarily suppressed in our model,
ensuring that our tree-level solutions survive radiative corrections.

%========================================================================
\section{The Lifetime of the Metastable Vacuum \label{sec:Lifetime}}

In order to be of phenomenological interest for model-building, 
the lifetime of any given metastable vacuum must be at least 
on the order of the present age of the universe.  
Such a state decays to the true vacuum via instanton transitions, 
the rate (per unit volume) for which may be parametrized as~\cite{Instantons}    
\begin{equation}
        \frac{\Gamma_{\rm inst}}{{\rm Vol}} ~=~ A\, e^{-B}~.
\end{equation}
We will not be particularly concerned with the form of the coefficient $A$, 
and will focus our attention on the exponent $B\equiv S_E(\phi)-S_E(\phi_+)$, 
usually referred to as the bounce action, which represents the difference between the
Euclidian action $S_E(\phi)$ at some point in field space 
and the action $S_E(\phi_+)$ at the metastable minimum.
In general, bounce actions must be either computed 
numerically or evaluated analytically in some limiting regime, since only a handful of 
potentials are known which admit exact, closed-form expressions for $B$.  

In this paper, we shall proceed as in Ref.~\cite{Triangles} by approximating the 
potential along the classical path between the metastable and stable vacua as a triangle. 
In this approximation, the bounce action depends 
on four parameters: $\Delta \phi_{\pm}$, which is the distance in field space between the top of the potential
barrier (in this case, Solution~C) and the metastable ($+$) or truly stable ($-$) vacuum (in this
case, Solutions~B and A respectively); 
and $\Delta V_{\pm}$, which is the potential difference between
the top of the barrier and each respective vacuum state.  
All four of these quantities are directly calculable 
in terms of $(\lambda,m,\xi_a,\xi_b)$
from the expressions in Eqs.~(\ref{solns}) and (\ref{Vs}).

Using the triangle approximation, calculating $B$ is relatively straightforward~\cite{Triangles}.
When 
\begin{equation}
  \frac{\Delta\phi_-}{\Delta\phi_+}~\geq~\frac{\sqrt{1+c}+1}{\sqrt{1+c}-1}~,
\label{eq:DecayIneq}
\end{equation}
with $c\equiv(\Delta V_-/\Delta V_+)(\Delta\phi_+/\Delta\phi_-)$, the bounce action 
is given by~\cite{Triangles}
\begin{equation}
  B ~=~\frac{32\pi^2}{3}\frac{1+c}{(\sqrt{1+c}-1)^4}\left(\frac{\Delta \phi_+^4}{\Delta V_+}\right)~.
\end{equation}
By contrast, when the inequality in Eq.~(\ref{eq:DecayIneq}) is not satisfied, 
the appropriate expression is instead given by~\cite{Triangles}
\begin{eqnarray}
   B ~&=&~\frac{\pi^2}{96}\left(\frac{\Delta V_+}{\Delta\phi_+}\right)^2 R_T^3\nonumber\\ 
        && ~~~~~ \times 
          \left(-\beta_+^3+3c\beta_+^2\beta_-+3c\beta_-^2\beta_+-c^2\beta_-^3\right)~,
\end{eqnarray}
where $\beta_{\pm}$ and $R_T$ are given by
\begin{equation}
     \beta_{\pm}\equiv \left(\frac{8\Delta \phi_{\pm}^2}{\Delta V_{\pm}}\right)^{1/2}~,~~~~~
       R_T\equiv \half\left(\frac{\beta_+^2+c\beta_-^2}{c\beta_- -\beta_+}\right)~.
\end{equation}
It can be verified that these solutions match smoothly at the point where Eq.~(\ref{eq:DecayIneq}) 
is saturated.

%================== FIGURE ============================================
\begin{figure}[ht]
\centerline{
   \epsfxsize 3.0 truein \epsfbox {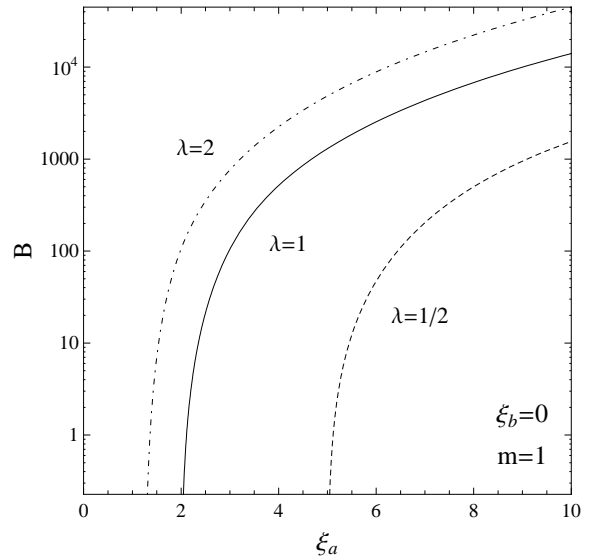} }
\smallskip
\caption{The bounce action 
       (i.e., the exponential suppression factor for the decay rate of our metastable vacuum),
    plotted as a function of $\xi_a$  
       for several different values of $\lambda$.
      For this plot, we have taken $m=1$ and $\xi_b=0$ (where the true ground state is
    supersymmetric).
    For sufficiently large $\xi_a$, we see that $B$ can easily exceed $\sim{\cal O}(10^3)$,
      implying that the lifetime of our metastable vacuum can easily exceed the age of the universe.}
\label{fig:RateSup}
\end{figure}
%================== END OF INSERTED FIGURE ============================

In Fig.~\ref{fig:RateSup}, we plot the bounce action $B$ as a function of $\xi_a$ for several different values of
$\lambda$.  For these plots, we have taken
$m=1$ and $\xi_b=0$ (implying that the true ground state preserves supersymmetry),
using the rescaled parameters introduced at the beginning of Sect.~III.~ 
Note that the bounce action $B$ --- and hence the lifetime of the false vacuum --- 
increases with increasing $\xi_a$.  Thus if $\xi_a$ is taken to be 
sufficiently large, the lifetime of the metastable vacuum in our model will exceed 
the present age of the universe, rendering such a vacuum phenomenologically viable.  
For smaller values of $\xi_a$, the lifetime of the metastable vacuum decreases until $\xi_a$ 
reaches its minimum allowed value in Eq.~(\ref{susycase}).  
At this point, the bounce action~$B$ 
drops to zero:  the metastable vacuum~B merges with 
the saddle-point Solution~C  and becomes unstable.

Note that with our original choice of model parameters $(\lambda,m,\xi_a,\xi_b)=(1,1,5,0)$ 
and $g=1$ as in Sect.~II, we find that $B\approx 1300$.  This is more than sufficient to guarantee
a metastable lifetime exceeding the age  of the universe  
even if the overall unspecified mass scale of our model is taken to be 
as high as the Planck scale.

%=============================================================
\section{Degenerate Ground States and a ``$\theta$-Vacuum''}

As we have seen, our model in Eq.~(\ref{eq:W}) gives rise
to metastable vacua.
However, as we shall now demonstrate, our model can also give rise
to something potentially even more interesting:
a doubly-degenerate classical ground state.
In a quantum-mechanical setting, this implies that our true
quantum-mechanical ground state is an effective ``$\theta$-vacuum''
of the theory.
This is yet another phenomenon illustrating the
highly non-trivial vacuum structure inherent in our model. 

To illustrate this phenomenon,
it is important to realize that in general, Solutions~A through E 
are not the only solutions to the equations in Eq.~(\ref{eq:ExistEqs}).  
While Solutions~A through E 
represent the complete set of solutions which exist along
the line $\xi_b= 0$, 
other solutions may also exist 
for other values of $(\xi_a,\xi_b)$. 
For example, one other solution giving rise to 
a stable vacuum is 
\beq
     F:~~~ v_3^2 = -\xi_b~,~~~~~ v_1^2=v_2^2=v_4^2=v_5^2=0~;
\eeq
this solution exists in the $\xi_b<0$ region, 
and is stable within the subregion 
\beq
          |\xi_a|<m^2~,~~~ |\xi_a| < -\lambda^2 \xi_b~.
\label{solnFstable}
\eeq
This vacuum even has unbroken supersymmetry when $\xi_a=0$.
Note that the regions of existence, stability, and unbroken
supersymmetry for Solution~F 
in the $(\xi_a,\xi_b)$ plane are the same as those for Solution~A,
only rotated by $\pi/2$ in the clockwise direction.

For $\lambda\leq 1$, the stable vacuum associated with Solution~F
does not simultaneously coexist with any other stable vacua.
However, for $\lambda>1$, the stable vacuum associated with Solution~F
coexists with the stable vacuum associated with Solution~A 
within the region
\beq
        \xi_b >-\lambda^2 \xi_a~,~~~ 
        \xi_b < \xi_a/\lambda^2 ~,~~~
        \xi_a,-\xi_b < m^2~.
\eeq
This then provides yet another region of parameter space in which
our model simultaneously contains both a true ground state 
and a metastable ground state, both non-supersymmetric.
For $\xi_a > -\xi_b$, the true ground state corresponds to Solution~A,
while for  
 $\xi_a < -\xi_b$, the true ground state corresponds to Solution~F.

An extremely interesting situation occurs when we further demand 
that $\xi_a= -\xi_b$:
along this line, 
our two stable solutions~A and F are exactly degenerate.  
This then provides an example of a situation in which
our model exhibits {\it two degenerate classical ground states}\/. 
Ultimately, 
the double degeneracy of the classical ground state
which emerges for $\xi_a= -\xi_b$ 
is the manifestation of  
a deeper $\IZ_2$ reflection  
symmetry
which involves 
the simultaneous exchanges of the two chiral superfields $\Phi_1\leftrightarrow \Phi_3$
and the two gauge groups $U(1)_a\leftrightarrow U(1)_b$.
In other words, this symmetry is a manifestation of the underlying reflection
(or ``parity'') symmetry of the original two-site moose upon which our model is built.  

Classically, a doubly degenerate vacuum state is an exciting
phenomenon.  Since the system must ultimately settle into one 
or the other ground state, what results is a spontaneous
breaking of the reflection/parity symmetry of our moose. 
One could even imagine the universe having chosen different
ground states in different spacetime locations, leading to 
domain-wall formation and other topological defects.

However, in a quantum-mechanical setting, the situation becomes
even more intriguing.  Because the scalar potential barrier
between the degenerate $|A\rangle$ and $|F\rangle$ vacua 
is finite, and because these two vacua are separated by only
a finite distance in field space, there will generally be tunneling between
these two vacua.  The existence of our moose reflection symmetry 
then implies that the true energy eigenstates of our system 
are not $|A\rangle$ or $|F\rangle$ independently, but rather
the reflection-symmetry eigenstates
\beq
       |\theta\rangle ~\equiv~ {1\over \sqrt{2}} 
               \left(  |A\rangle + e^{i\theta} |F\rangle \right)~~~~
            {\rm for}~~~\theta\in\lbrace 0,\pi\rbrace~.
\eeq
This is exactly what we would expect from 
an ordinary quantum-mechanical double-well analysis.
As a result of the mixing between these states, the energy of the
symmetric $|\theta{=}0\rangle$ state is lower than that of the
anti-symmetric $|\theta{=}\pi\rangle$ state. 
As a result, the true quantum-mechanical vacuum of our theory
is the $|\theta{=}0\rangle$ state,
while the 
$|\theta{=}\pi\rangle$ state becomes a metastable vacuum.
This, then, provides yet another way of achieving metastability 
in our model.  Note that the energy of the true ground state
in such a setup is lower than the classical energy associated   
with either Solution~A or Solution~F individually.
Of course, despite our use of the $\theta$-variable, 
we emphasize that this sort of ``$\theta$-vacuum''
should not be confused with the traditional QCD $\theta$-vacuum;
indeed, the only gauge symmetries in our model are abelian.

It is interesting to speculate that in a more complicated model,
even more degenerate ground states might emerge.
The true vacua of such a model  would then correspond to Bloch waves
across these degenerate ground states, with the corresponding energy eigenvalues populating
nearly continuous energy ``bands''.  This could then provide one possible way
of realizing the proposal in Ref.~\cite{Gordy} for addressing
the cosmological-constant problem.

%========================================================================
\section{Discussion\label{sec:Discussion}}

In this paper, we have presented a remarkably simple construction which simultaneously 
gives rise to a supersymmetric ground state and a non-supersymmetric metastable
state, with both arising classically in a theory with no flat directions or
infinite distances in field space.
The key feature of our construction is that the supersymmetry-breaking 
in our model arises not only through $F$-term breaking, but also through $D$-term breaking.
Since all of the relevant physics is perturbative, 
we were able to perform
explicit calculations of the lifetimes and particle spectra associated with such vacua
and demonstrate that these lifetimes can easily 
exceed the present age of the universe.   

As we have shown, 
the supersymmetry and R-symmetry in our construction are broken at tree level in a 
perturbative theory where no flat directions appear in 
the classical potential and where all minima appear at finite locations 
in field space.
This gives our construction a distinct advantage
relative to constructions which have appeared in much of the prior 
literature on metastable supersymmetry-breaking.
For example, the effective O'Raifeartaigh-inspired models discussed in Refs.~\cite{ISSR,Shih} 
contain runaway directions along which supersymmetric global minima occur 
at infinite distances in field space.  While it has been demonstrated~\cite{ISS} 
that runaway directions of this sort can be successfully stabilized 
by non-perturbative dynamics at large but finite field values, the presence of such 
non-perturbative dynamics renders the potential barrier between stable and metastable vacua 
difficult to describe.  As a result, explicit lifetime 
calculations are challenging to perform.              

The prior literature also contains 
models in which metastable supersymmetry-breaking occurs at finite field values.
However, these scenarios generally contain classical flat directions which 
may or may not be lifted at the quantum level.  For example, the model 
described in Ref.~\cite{EllisEtAl} leads to both supersymmetric and non-supersymmetric 
flat directions at tree level, separated by a flat-topped ``ridge'' of degenerate 
points in field space.  Although the degeneracy along the non-supersymmetric flat direction 
(as well as that along the ridge) is lifted at the quantum level, the 
supersymmetric flat direction remains flat to all orders in perturbation theory.  
Thus, one would require non-perturbative physics or some modification of the
model itself in order to achieve a truly stable ground state.
Likewise, the model outlined in Ref.~\cite{Benakli} also contains a metastable flat 
direction at tree level.  Since this flat direction is destabilized by quantum corrections, 
some additional dynamics (such as a supergravity contribution 
to the scalar potential, as discussed by the authors of Ref.~\cite{Benakli}) 
is required for true stability.  

By contrast, the ``kernel'' presented in this work 
utilizes only finite field values and perturbative dynamics, yet
possesses neither classical flat 
directions nor runaway behavior.
As such, we believe that this represents an interesting alternative to 
previous supersymmetry-breaking scenarios. 
Indeed, our experience suggests that within the context of traditional
dynamical supersymmetry-breaking scenarios, such models are relatively rare, 
and we are not aware of any models with these 
characteristics in the prior metastability literature.
Moreover, while the supersymmetry-breaking in our model is sourced by $D$-terms,
this in turn triggers $F$-term breaking as well.
Thus our kernel avoids the phenomenological problems normally associated
with pure $D$-term-breaking models, and indeed
a potential method of mediating supersymmetry-breaking to a visible
sector was outlined below Eq.~(\ref{extravecs}).

Needless to say, 
metastability in and of itself is not a phenomenological necessity.
Indeed, there already exist models in the literature~\cite{UVcompletion}
in which supersymmetry and R-symmetry are broken at tree level in the true ground state
and in which no metastable vacua appear.
However, we believe that metastability offers rich possibilities when it 
opens up new models to phenomenological viability
that would previously have been deemed problematic based on the properties
of their ground states.
This is certainly true in the model of Ref.~\cite{ISS} as well as other similar models 
it has inspired, since the ground states of such models are supersymmetric and hence
unacceptable on phenomenological grounds.  However, 
this is also true in models with
multiple $U(1)$ gauge
groups (which form the context of the discussion of this paper) 
because the ground states of such models
frequently involve pure $D$-term breaking --- a situation which also yields
phenomenological difficulties --- as well as potentially unbroken supersymmetry.
Thus, metastable supersymmetry-breaking offers a phenomenological advantage
within such models as well, and suggests that
our  model may have broad implications for an even wider
class of theories than we have considered here.
Moreover, as we pointed out,
such models can play the role of self-sufficient supersymmetry-breaking ``kernels''  
which are not overly sensitive to the larger models in which they are embedded.
As a result, their supersymmetry-breaking properties  are largely unaffected
by the additional physics that would be required in order to build a complete
phenomenological model of supersymmetry-breaking.

Strong motivation to consider such models and their phenomenology
also comes from the fact that
structures involving additional $U(1)$ gauge groups arise with great frequency in string theory.
Indeed, from this perspective, our work can be interpreted as highlighting an important fact about models involving
Fayet-Iliopoulos terms and about the landscape of string-motivated
models in general, indicating that substantial regions of the landscape that would not have
previously been considered phenomenologically viable are in fact so.      
This is illustrated by the explicit examinations of the parameter space 
that we performed in Sect.~III.

Before we conclude, a few additional comments are in order.  
First, the kernel presented in Sect.~II includes two dimensionful 
parameters, $\sqrt{\xi_a}$ and $m$.  While it is beyond the
scope of this paper to present a specific mechanism for dynamically 
generating these parameters at scales parametrically smaller
than the Planck scale $M_{\rm Planck}$, we will now argue that arranging such a separation of scales 
may well be feasible.
We mentioned in Sect.~II that this 
kernel naturally appears in the low-energy limit of heterotic string models. 
In such a context, the variant~\cite{DineSeibergWitten} of the Green-Schwarz mechanism~\cite{GreenSchwarz} 
that can be used to cancel the mixed gauge anomalies implicit in the charge assignments given 
in Table~\ref{tab:chgs} also gives rise to the Fayet-Iliopoulos terms 
that break supersymmetry.  However, it has been argued~\cite{DudasDecon} that the physical scale 
associated with these terms is not tied to the Planck scale and therefore the supersymmetry-breaking 
scale can be much smaller than $M_{\rm Planck}$.  Alternatively, one could imagine
generating $\xi_a$, $\xi_b$, and $m^2$ dynamically via the retrofitting methods of Ref.~\cite{DineRetro},
or generating an ultraviolet completion of our model~\cite{Future}, perhaps 
using D-brane configurations
along the lines developed in Ref.~\cite{UVcompletion}.

As a result of the robustness of the metastable vacuum in our model against small corrections,
most of the standard methods of mediating supersymmetry-breaking to the visible sector 
(gravity mediation~\cite{GravMed}, gauge 
mediation~\cite{DineNelsonMetastable,PreISSDine,GaugeMed}, etc.)\  should be viable options 
for our model.
Indeed, we saw in Sect.~III that introducing additional vector-like pairs 
of chiral superfields with gauge charges identical to those of $\Phi_4$ and $\Phi_5$ 
and supersymmetric masses greater than $m$ does not destabilize the 
vacuum structure of the theory.  
This, therefore, suggests a
simple method of coupling our kernel to a messenger sector.  In addition, 
the presence of multiple $U(1)$ groups makes $Z'$ mediation~\cite{ZPrimeMed} an intriguing 
alternative possibility, and perhaps the most natural way of mediating supersymmetry-breaking
to the fields of the minimal supersymmetric standard model (MSSM).  A scenario of this sort, 
in which some of the MSSM fields would be charged under one or more hidden-sector 
$U(1)$ groups, could lead to a rich $Z'$ phenomenology potentially visible at 
the LHC~\cite{Future}.
It would also be interesting to incorporate kinetic-mixing effects~\cite{kinetic} into our analysis~\cite{Future}.

The cosmological implications of this construction are also quite interesting.  
As discussed in Sect.~V, regions of parameter space in this scenario which preserve the
$\IZ_2$ reflection symmetry of the two-site moose give rise to a two-fold vacuum 
degeneracy and a ground state with vacuum energy lower than that of either of the degenerate 
classical solutions alone.  One could also imagine similar scenarios~\cite{Future} where a far larger number of
degenerate or nearly degenerate vacua exist, resulting in a band structure of the 
kind explored in Ref.~\cite{Gordy}.  It may also be possible to construct an extension of 
this setup in which the cosmological-constant problem is addressed via transitions through a 
set of non-degenerate vacua, as in Ref.~\cite{BoussoPolchinski}, or through other non-traditional
tunnelling effects~\cite{tye} in a vast stringy cosmic landscape.  
The presence of additional $U(1)$ gauge groups in this setup also has implications 
for cosmology, as they will inevitably give rise to
cosmic strings~\cite{CosmicStrings} and other topological defects with rich
phenomenologies~\cite{CosmicStringApps}.
 
Finally, as already mentioned above, 
the presence of long-lived metastable 
vacua and non-trivial vacuum structures 
can have a significant impact on state counting
in studies of the string-theory landscape. 
Because additional $U(1)$ groups with non-trivial 
Fayet-Iliopoulos terms are a general feature of heterotic and Type~I string models, 
our results suggest that
metastable vacua (many of which break supersymmetry, and
at potentially high scales) exist on a substantial fraction of the string landscape and must
therefore be taken into account when statistical surveys of the landscape are performed.  
To do this properly, one would have to account for vacuum transitions at finite temperature and
ascertain the thermal population of the landscape, taking into account the fact that ``holes'' 
in the landscape (i.e., regions where no stable vacuum exists) have been shown to occur in 
deconstruction-motivated models~\cite{DDG} and can result in isolated
``islands''~\cite{Islands} that are not in thermal contact with one another.

We see, then, that the ``kernel'' we have presented in this paper may be of interest 
for constructing models of low-scale supersymmetry-breaking, for understanding
the phenomenology of $Z'$ physics, and for evaluating various properties of the  
string landscape.
As such, we expect that these sorts of models can serve as fertile arenas from which
future investigations into the behavior of supersymmetric field theories might hatch.

%   {\tt
%  \begin{verse}
%  
%  So much depends\\
%  upon \\
%  \bigskip
%  the vacuum\\
%  structure \\
%  \bigskip
%  glazed with\\
%  excitations\\
%  \bigskip
%  forming white\\
%  chickens.~\cite{Wheelbarrow}\\
%  
%  
%  \end{verse}
%  }

%============================================================================= 
\section*{Acknowledgments}

This work was supported in part
by the Department of Energy under Grant~DE-FG02-04ER-41298.
We wish to thank
J.~Bourjaily, Z.~Chacko, E.~Dudas, and M.~Lennek
for discussions.
%  and R.~Wagner and B.~Dylan for inspiration concerning
%  the importance of the ground state and metastable states respectively.

%=============================================================================

\medskip

%========================================================================
%========================================================================
%========================================================================

\end{document}